\documentclass[12pt]{iopart}
\usepackage{graphicx}
\usepackage{epsfig}
\begin{document}

\title[Direct visualization of aging...]{Direct
visualization of aging in colloidal glasses}

\author{Rachel E.~Courtland\ and Eric R.~Weeks
\footnote[1]{To
whom correspondence should be addressed (weeks@physics.emory.edu)}
}

\address{Physics Department, Emory University, Atlanta, GA 30322
USA}

\begin{abstract}
We use confocal microscopy to directly visualize the dynamics of
aging colloidal glasses.  We prepare a colloidal suspension at
high density, a simple model system which shares many properties
with other glasses, and initiate experiments by 
stirring the sample.  We follow the motion of several thousand
colloidal particles after the stirring and observe that their
motion significantly slows as the sample ages.  The
aging is both spatially and temporally heterogeneous.
Furthermore, while the characteristic relaxation time scale
grows with the age of the sample, nontrivial particle motions
continue to occur on all time scales.

\end{abstract}


\submitto{\JPCM}


\section{Introduction}

Colloidal suspensions have long been used
as model systems to study the glass transition
\cite{pusey,bartsch93,vanmegen94,alfons95,reviews}.  As the
concentration of a colloidal suspension is increased, the
motion of the colloidal particles becomes increasingly slowed.
The glass transition is considered to occur at the concentration
at which particle motion ceases to be diffusive on long time
scales \cite{reviews}.  Recent work has studied the behavior
of simple colloidal suspensions as this transition is approached
\cite{weeksweitz,weekscrocker,kegel00}.  In this paper we
examine the behavior of glassy samples.  Unlike supercooled
colloidal fluids,
the behavior of a colloidal glass depends on
the history of the sample: particle motion becomes increasingly
slowed as the sample ages \cite{bouchaud,vanmegen98,luca00,bonn98,bonn00}.
We use confocal microscopy to study the three-dimensional
motions of colloidal particles in an aging colloidal glass.
We find that aging occurs due to slight rearrangements which
are both spatially and temporally heterogeneous.  Moreover, even
after the sample has aged for long times, small but significant
motions occur on short time scales.  Previous work has seen
similar evidence of spatial and temporal heterogeneities in
aging systems \cite{bouchaud,katerina02,macphail97,castillo02}.

\section{Experimental Methods}

We use poly-(methylmethacrylate) (PMMA) particles of radius
$a = 1.18$ $\mu$m and polydispersity $\sim5 \%$, sterically
stabilized by a thin layer of poly-12-hydroxystearic acid.
The particles are dyed with rhodamine and suspended in a
mixture of organic solvents (cyclohexylbromide and decalin)
that closely matches both the density and index of refraction
of the particles.  We use a scanning laser confocal microscope
to acquire images of a viewing volume of 63 $\mu$m $\times$ 58
$\mu$m $\times$ 12 $\mu$m at the rate of 3 images per minute.
The viewing volume typically contains $\sim$2400 particles.
We focus at least 60 $\mu$m away from the cover slip of the
sample chamber to avoid wall effects.  We identify particles
with a horizontal accuracy of 0.03 $\mu$m and a vertical
accuracy of 0.05 $\mu$m and track them in three dimensions
over the course of the experiment \cite{dinsmore01,crocker96}.

The control parameter for the colloidal phase behavior is
the sample volume fraction $\phi$.  While the rhodamine
imparts a slight charge upon the particles, their phase
behavior, $\phi_{freeze} = 0.38$ and $\phi_{melt} = 0.42$,
is similar to that of hard spheres ($\phi_{freeze} = 0.494$
and $\phi_{melt} = 0.545$).  We observe a glass transition
at $\phi_g \approx 0.58$, in agreement with what is seen for
hard spheres \cite{pusey,kegel00}.  We examined samples with volume
fractions ranging $\phi \approx$ 0.58 to $\phi \approx$ 0.62.
These samples form small crystals which nucleate at the
coverslip, but do not form crystals within the bulk of
the sample even after several weeks.

In order to initialize the system, a small length of wire is
inserted into each sample chamber.  After placing a sample
on the microscope stage, a handheld magnet is used to pull
and rotate the wire through the sample for several minutes.
The subsequent particle dynamics are reproducible after this
stirring.  Within a minute of ending the stirring, transient
flows within the sample greatly diminish, and the particles
move slowly enough to be identified and tracked.  This
defined our initial time $t=0$ for each sample, although the
results below are not sensitive to variations of this choice.
During the course of the experiments, no crystallization was
observed within the viewing volume.  Note that in many aging
studies, the initial sample is prepared by a temperature quench
(which would correspond to a rapid increase of the volume
fraction in our experiments).  The initial conditions in our
experiments correspond to a shear-melted sample; the volume
fraction remains constant.

\section{Results}

\begin{figure}
\begin{center}
\epsfxsize=7.0truecm
\epsffile{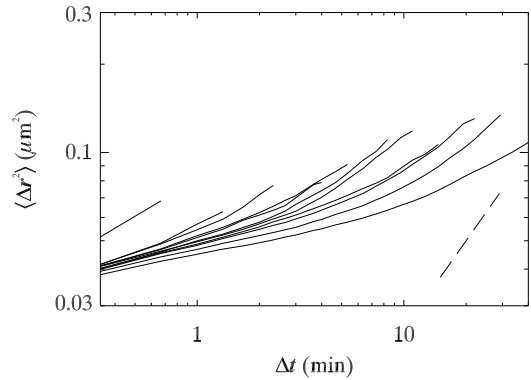}
\caption{\label{fig1}Mean square displacement versus lag time for a colloidal 
glass ($\phi \approx 0.62$).
The total length of the data set is 120 min.  Data is divided 
logarithmically according to sample age.  The
geometric mean ages of the curves, from 
left to right, are $\overline{t_w} =$ 
3.0, 4.2, 5.9, 8.2, 12, 16, 23, 32, 44, 62, 87 min.
The dashed line has a
slope of 1.}
\end{center}
\end{figure}

As a glass ages, we expect dynamical properties to change
with time.  To investigate this, we consider small temporal
portions of a data set and calculate the mean square
displacement $\langle \Delta r^2 \rangle$ (MSD) for each
portion, where the angle brackets indicate an average over
all particles and all initial times within each temporal
portion.  Figure~1 displays the progression of the MSD curves
of a colloidal glass ($\phi=0.62$) over a period of two hours, showing
a marked change as the sample ages.  The data is divided
logarithmically into segments according to the sample age
$t_w$,  the time waited since stirring, such that $t_w^{n+1}
= c t_w^n$, where $c = 1.4$, $t_w^0 = 3.0$ minutes and $n =
[1, 2, ... ,11]$.  The MSD for a given $t_w$ is calculated
using data within the interval [$c^{-1}t_w$, $c t_w$].
The MSD behavior at low and
intermediate lag times is consistent with that of super-cooled
liquids \cite{weeksweitz,weekscrocker}.  For $\Delta t < 30$~s
(not shown), the MSD grows roughly
linearly with $\Delta t$, corresponding to the diffusion of
particles inside ``cages'' formed by neighboring particles
\cite{reviews,weeksweitz}.  The MSD at intermediate lag times
exhibits a plateau, as particles are locally confined by their
neighbors.  At larger lag times, the MSD curves each show a
slight upturn, reminiscent of super-cooled liquids \cite{bonn00}.
For super-cooled liquids, this upturn corresponds to cage
rearrangements:  the configuration of particles around a
caged particle changes noticeably \cite{weeksweitz}.  In our
aging samples, we see similar, though smaller, motions,
and the upturn in the MSD is less obvious than is seen in liquids.
These motions, which change the local positions of particles,
are likely related to cage rearrangements \cite{bonn98,bonn00}.
Presumably these local rearrangements lead to the
aging of the sample.  However, unlike super-cooled liquids,
the time scale for the upturn in the MSD depends strongly
on the age of the sample, $t_w$, as is clear from Fig.~1
\cite{bonn00}.
For larger values of $t_w$, the plateau extends over a larger
range of $\Delta t$.  This behavior has been seen before in
a variety of aging systems \cite{vanmegen98,luca00}.

We observe also that the MSD of a colloidal glass does not
evolve uniformly.  Consider the four rightmost curves in
Fig.~1, corresponding to $\overline{t_w}=$ 32, 44, 62, and
87 min.  The first two curves ($\overline{t_w}$=32, 44 min)
are virtually indistinguishable, showing that the dynamics
have not changed within the imaging volume.  However, the next
two curves ($\overline{t_w}$=62, 87 min) are quite different,
showing that the particles within the imaging volume have aged.
This suggests that aging is in part temporally heterogeneous
in our small observation volume.  It is to be expected that if a
larger volume was examined, such as in light scattering
experiments, the dynamics would appear to evolve more smoothly
with $t_w$ \cite{vanmegen98,luca00}.

\begin{figure}
\begin{center}
\epsfxsize=6.0truecm
\epsffile{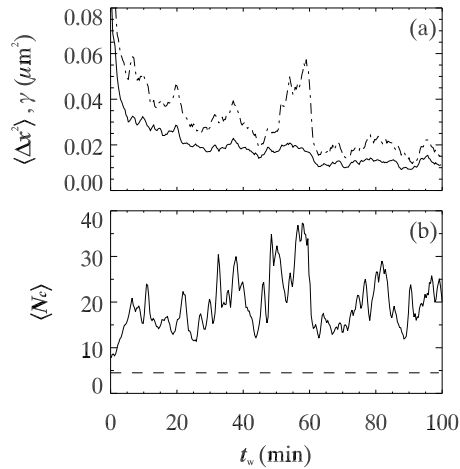}
\caption{\label{fig2}Various ensemble features as a function of sample   
age, $t_w$.  The time lag is fixed ($\Delta T$ = 10 min) and the
data are the same as shown in Fig.~1.  (a) 
$\langle \Delta x^2 \rangle$ and 
$\gamma = \sqrt{\langle \Delta x^4 \rangle /3}$
(dashed).  (The $x$ component is used as the $z$ component is
noisier due to optical effects.)
(b) Average number of particles in a cluster of mobile
particles; see text for details.  The dashed line indicates the
average cluster size if mobile particles were randomly
distributed.
}
\end{center}
\end{figure}

The temporal averaging used in calculating the MSD obscures
some of the details of this temporal heterogeneity.  However,
it is unclear how to best study the dynamics given that the time
scales for such dynamics may vary dramatically with the sample
age $t_w$.  Furthermore, the mobile particles responsible for
the upturn in the MSD are only moving slightly farther than
the immobile particles;  the latter
still move locally due to their Brownian motion.
To help distinguish the rearranging particles from those
which only move within their cage, we average the trajectory
of each particle over 1 min.  The results that follow are not
sensitive to the choice of this averaging time.

To investigate the nature of the temporally
heterogeneous aging, we choose a fixed lag time $\Delta T$
and study how the dynamical behavior changes with $t_w$.
We might expect that for $t_w \ll \Delta T$, $t_w \approx
\Delta T$, and $t_w \gg \Delta T$, we should see strikingly
different behavior.  We choose
$\Delta T$ = 10 min to allow
us to resolve these three regimes with our data.  The solid
line in Figure 2(a) shows $\langle \Delta x^2 \rangle_{\Delta T}$
plotted as a function of sample age, $t_w$; here, the angle
brackets do not indicate a time average, but only a particle
average, unlike the MSD.  As $t_w$ increases, $\langle \Delta
x^2 \rangle$ generally decreases, which is not surprising given
the behavior shown in Fig.~1:  at large $t_w$, particle motion
on a time scale of $\Delta T$=10 min reflects the motion of
particles confined by their neighbors.  However, in addition
to this overall relaxation of $\langle \Delta x^2 \rangle$,
Fig.~2(a) shows a number of abrupt events, when $\langle
\Delta x^2 \rangle$ jumps in value before relaxing further.
Further confirmation of the temporally heterogeneous nature of
this process is demonstrated by the dashed line in Fig.~2(a),
which shows $\gamma = \sqrt{\langle \Delta x^4 \rangle /3}$.  If the
dynamics were purely gaussian, i.e. diffusive, we would expect
$\gamma=\langle \Delta x^2 \rangle$.  The consistently higher
value of $\gamma$ and its prominent correlation to events
in $\langle \Delta x^2 \rangle$ indicate that the ensemble
dynamics are highly nongaussian and events in $\langle \Delta
x^2 \rangle$ particularly so.

\begin{figure}
\begin{center}
\epsfxsize=11.0truecm
\epsffile{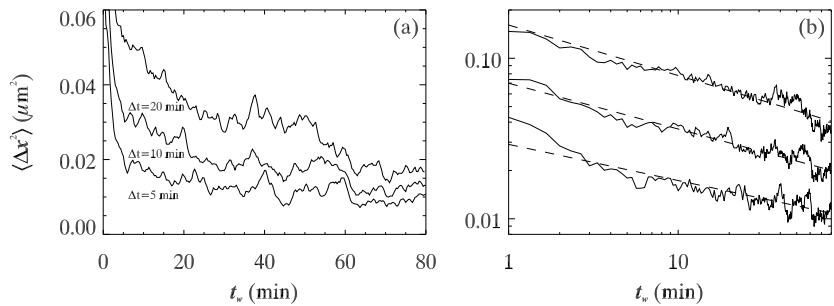}
\caption{\label{fig3}$\langle \Delta x^2 \rangle$ as a function of
sample age for various values of lag time, $\Delta t$, on linear
axes (a) and log-log axes (b).  
The straight lines in (b) are fits to 
power law decay, $\langle \Delta r^2 \rangle \sim {t_w}^{-a}$,
with $a$=0.23, 0.29, and 0.31 for $\Delta t$ = 5, 10, and 20 min,
respectively.  The curves in (b)
corresponding to $\Delta t$ = 5 and 20 minutes 
have been vertically offset for clarity.  
The data are the same as in the previous 
figures. }
\end{center}
\end{figure}

It is intriguing that there is no unique behavior for $t_w \approx
\Delta T$; instead, aging ``events" occur at many different
values of $t_w$.  To further confirm this, we plot $\langle
\Delta x^2 \rangle$ versus sample age for several different
values of $\Delta T$ in Fig.~3.  The curves are clearly related;
many events are simultaneously present in all curves.  In Fig.~3(b) these
curves are plotted on log-log axes, showing that the overall
decrease in $\langle \Delta x^2 \rangle$ is consistent with a
power law decay, $\langle \Delta x^2 \rangle \sim {t_w}^{-a}$.
For different samples and different choices of $\Delta T$,
we find $a$ ranges between 0.05 and 0.5.

In super-cooled colloidal liquids it is known that the dynamics
are spatially heterogeneous as well as temporally heterogeneous
\cite{weekscrocker,kegel00}.  To investigate this for our
aging samples, we plot the 3D locations of mobile particles
in Fig.~4 for $t_w=$ 10, 55, and 95 min.  The mobility of a
particle is defined as its displacement during the 
interval $t_w$ to $t_w + \Delta T$ (with $\Delta t =10$ min).
The particles shown are the
10\% most mobile at those times, and are generally grouped
into large clusters.  Varying thresholds does not
substantially change these pictures.  Our results are similar
to the spatial heterogeneities seen in simulations of aging
\cite{katerina02}.  Signs of heterogeneities have also been seen
in aging experiments which studied glycerol \cite{macphail97}.
Previous microscopic studies of colloidal glasses did not
show such large clusters \cite{weekscrocker}; we find that
the slight averaging of the particle trajectories discussed
above (over a 1 min interval) is necessary to distinguish
the rearrangements from the local Brownian motion of caged
particles.  Without averaging, our pictures look similar
to those of Ref.~\cite{weekscrocker}.

It is striking that the mobile particles are clustered for all
values of $t_w$, notably at both $t_w = 10$ min $=\Delta T$
and $t_w = 95$ min $\gg \Delta T$.  We investigate the $t_w$
dependence of the spatial clustering of mobile particles by
generating similar pictures for each $t_w$ and measuring
the average size of the clusters in each picture.  (A cluster
is defined as all connected mobile particles which are closer
to each other than 3.2 $\mu$m, the first minimum of $g(r)$,
the pair correlation function.)  Again, mobile particles are
defined as the particles with displacements in the top 10\% at
each $t_w$, using $\Delta T$ to characterize the displacements.
The average cluster size shows large fluctuations, as seen
in Fig.~2(b).  If the 10\% most mobile particles were randomly
distributed in space, the average cluster size would be
$\sim 4.5$ [dashed line in Fig.~2(b)], however, we find that
the average cluster size is always larger than this value.
The larger average cluster sizes correspond to locally large
values of $\langle \Delta x^2 \rangle$ and $\gamma$, seen
in Fig.~2(a).

\begin{figure}
\begin{center}
\epsfxsize=5.0truecm
\epsffile{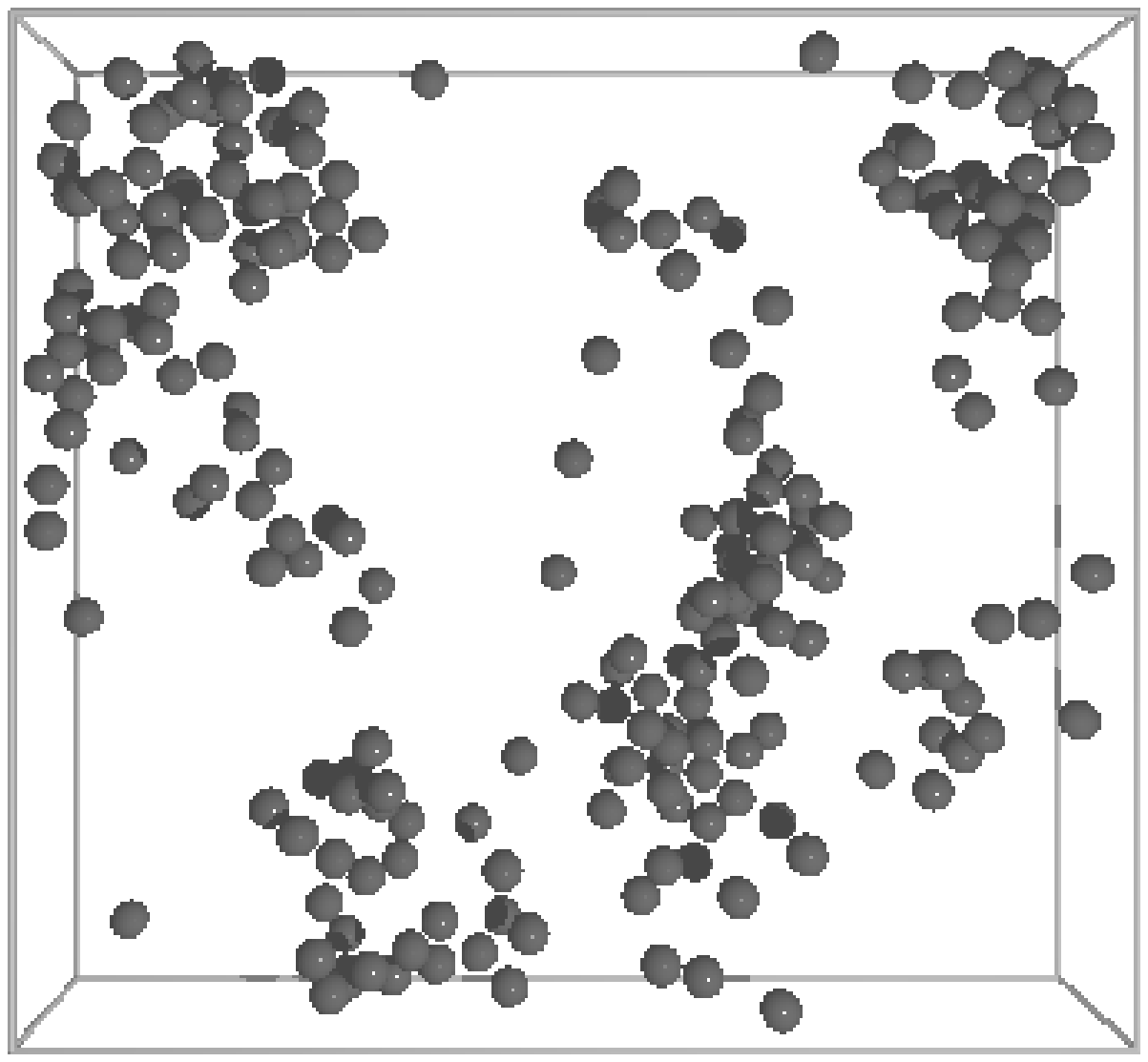}
\epsfxsize=5.0truecm
\epsffile{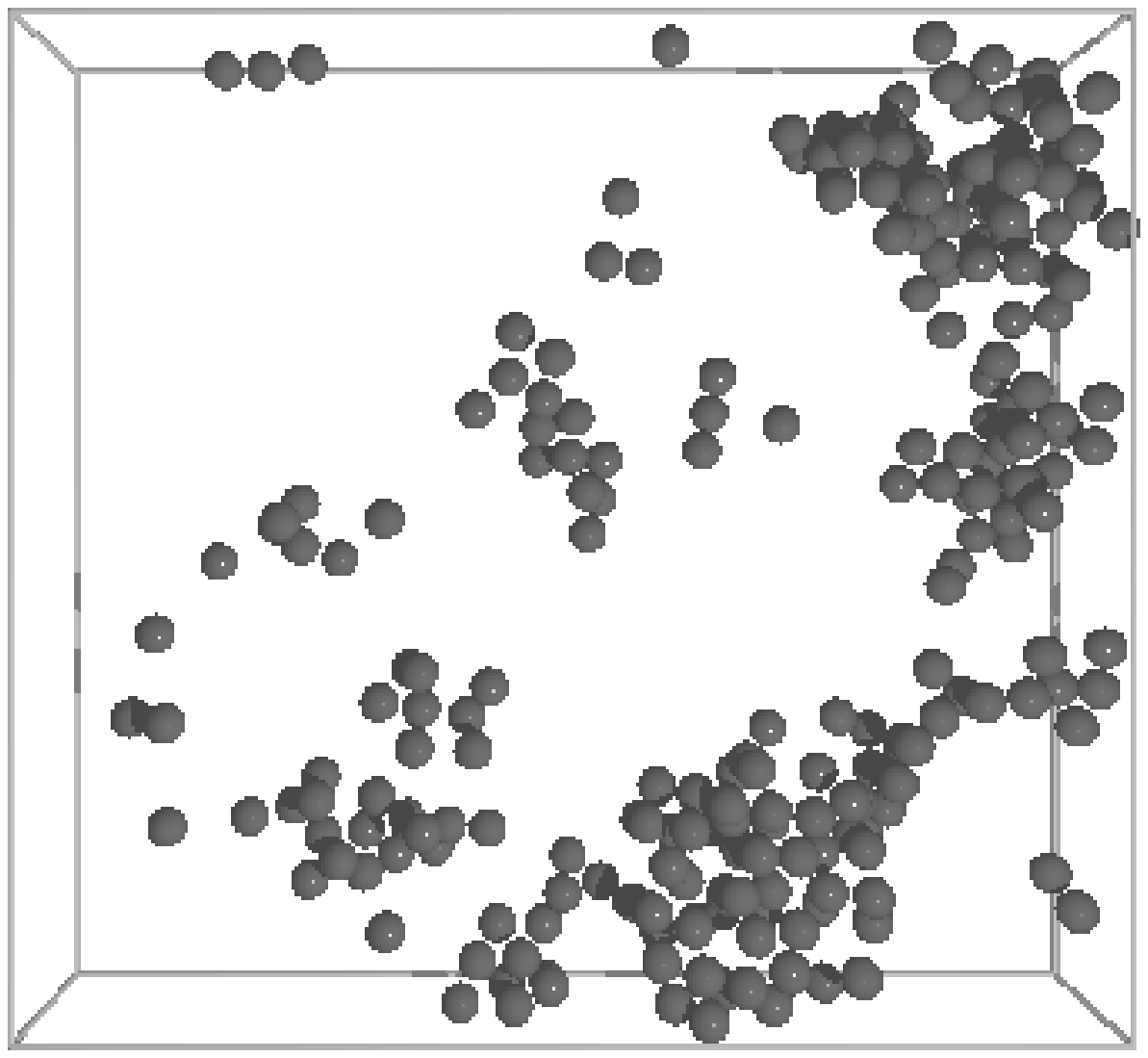}
\epsfxsize=5.0truecm
\epsffile{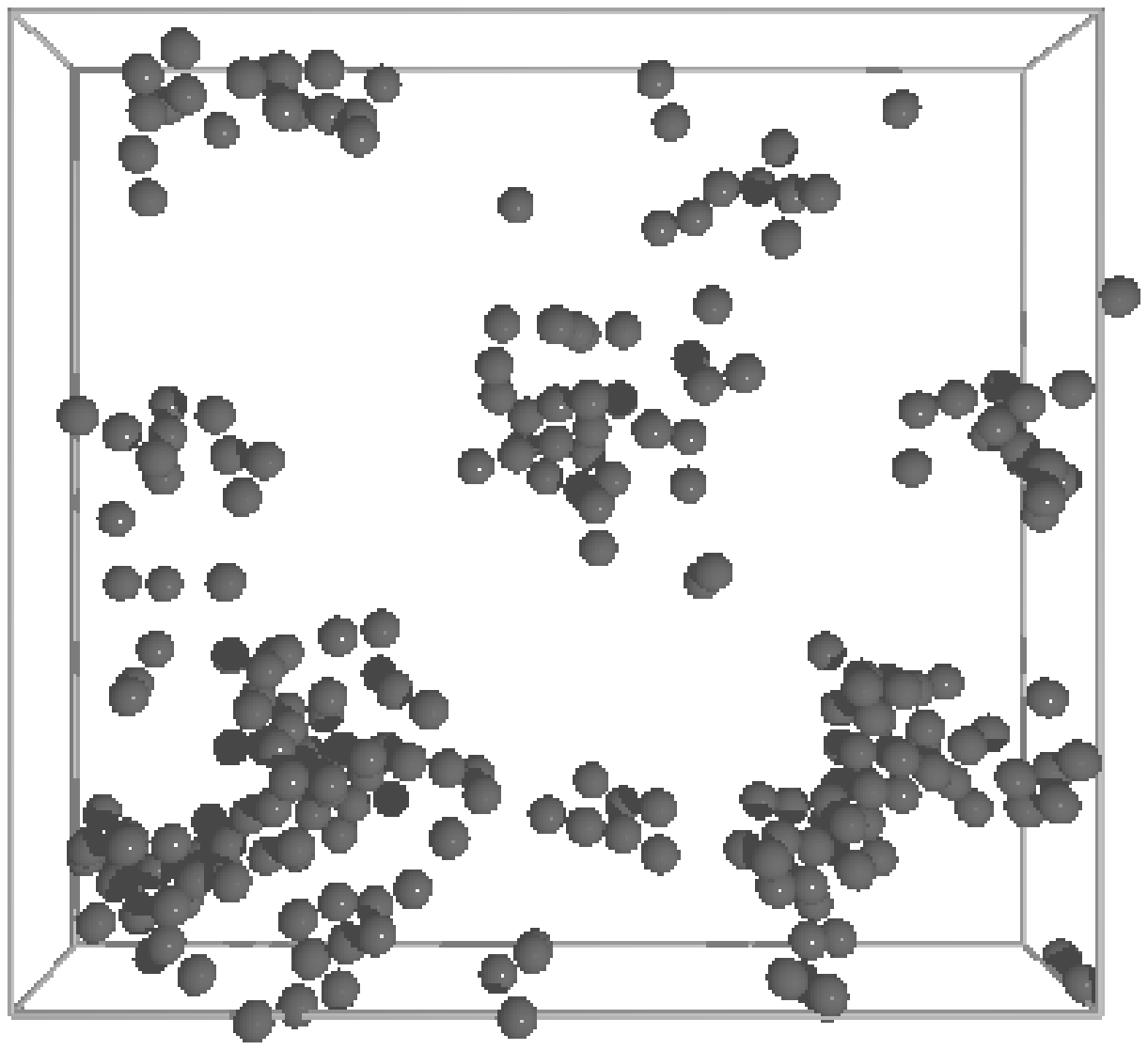}
\caption{\label{clusters}
Locations of the 10\% most mobile particles at three
different ages $t_w$.
For each picture, mobility was determined by calculating
displacements $\Delta r$ over an interval $[t_w,t_w + \Delta T]$,
with $\Delta T=10$ min.
Left: $t_w = 10$ min, and $\Delta r > 0.43$ $\mu$m for the most
mobile particles.
Middle: $t_w = 55$ min, $\Delta r > 0.34$ $\mu$m.
Right: $t_w = 95$ min, $\Delta r > 0.33$ $\mu$m.
The data are the same as shown in previous figures, and the
choices of $t_w$ correspond to local maxima of $\gamma$ in
Fig.~2(a).  The particles are drawn to scale (2.36~$\mu$m
diameter) and the box shown is the entire viewing volume (within a much
larger sample chamber).
}
\end{center}
\end{figure}

Our most surprising result is that overall, the cluster
sizes do not increase dramatically with $t_w$.  For the
sample to age, particles must move, and may do so in small
groups such as these clusters.  For early ages $t_w$, it
might be expected that particle rearrangements may involve
very few particles, which can rearrange relatively quickly,
thus causing the MSD to have a shorter plateau.  For larger
$t_w$, if the rearrangements involved more particles, this
would by necessity be more difficult and take longer to occur,
thus explaining the slower dynamics reflected in the later MSD
curves of Fig.~1.  However, this is not what we see.  As seen
in Fig.~2(b), the average cluster size does not dramatically
change as a function of $t_w$.  This remains true even if we
scale the choice of $\Delta t$ with $t_w$ (not shown).  Thus,
it is not yet clear what ultimately causes the dynamics to slow
as our sample ages.  While the rearrangements occur via groups
such as those shown in Fig.~4, we have not found anything in
their character which changes as the sample ages.

While all the data shown in the figures correspond to one
sample with $\phi = 0.62$, we have investigated five other
samples in the range $0.58 \leq \phi \leq 0.62$.  All data
sets show similar behavior, and as yet we see no clear $\phi$
dependence in our results.

\section{Conclusions}

Aging in our colloidal samples appears to be due to the most
mobile particles in the sample, seen in Fig.~4 and reflected by
the fluctuations in Fig.~2.  In order for the sample to age,
particles must move to new positions.  Our results show
that such rearrangements are both spatially and temporally
heterogeneous.  In general, each aging event appears spatially
uncorrelated with the previous.  The surprising result is
that significant fluctuations occur on time scales $\Delta
t \ll t_w$, despite the fact that for large $t_w$, the mean
square displacement shows only a plateau at such $\Delta t$
(see Fig.~1).  We find, in fact, that the plateau in the
MSD is due to a temporal average over these infrequent
rearrangement events and the more prevalent caged behavior.
We have also attempted to investigate the behavior of the
most immobile particles, but this has been difficult, as
every particle still possesses nontrivial Brownian motion,
and there is no unambiguous way to distinguish the most
immobile particles.

Thus, the changing character of the MSD (Fig.~1) and the
related decay in $\langle \Delta x^2 \rangle$ for a fixed $\Delta t$
(shown in Fig.~3) seem to be the most significant dynamical
changes in aging colloidal glasses.  In our future work,
we plan to investigate the microscopic structural changes
associated with these dynamical changes.

\section{Acknowledgments}

We thank J.-P.~Bouchaud and S.~Franklin for helpful discussions.
This work was supported by NASA and the University Research
Committee of Emory University.

\end{document}